# Deep Learning Approach for Dynamic Sampling for Multichannel Mass Spectrometry Imaging

David Helminiak, *Member, IEEE*, Hang Hu, Julia Laskin, and Dong Hye Ye, *Member, IEEE*

*Abstract*—Mass Spectrometry Imaging (MSI), using traditional rectilinear scanning, takes hours to days for high spatial resolution acquisitions. Given that most pixels within a sample's field of view are often neither relevant to underlying biological structures nor chemically informative, MSI presents as a prime candidate for integration with sparse and dynamic sampling algorithms. During a scan, stochastic models determine which locations probabilistically contain information critical to the generation of low-error reconstructions. Decreasing the number of required physical measurements thereby minimizes overall acquisition times. A Deep Learning Approach for Dynamic Sampling (DLADS), utilizing a Convolutional Neural Network (CNN) and encapsulating molecular mass intensity distributions within a third dimension, demonstrates a simulated 70% throughput improvement for Nanospray Desorption Electrospray Ionization (nano-DESI) MSI tissues. Evaluations are conducted between DLADS and a Supervised Learning Approach for Dynamic Sampling, with Least-Squares regression (SLADS-LS) and a Multi-Layer Perceptron (MLP) network (SLADS-Net). When compared with SLADS-LS, limited to a single $m/z$ channel, as well as multichannel SLADS-LS and SLADS-Net, DLADS respectively improves regression performance by 36.7%, 7.0%, and 6.2%, resulting in gains to reconstruction quality of 6.0%, 2.1%, and 3.4% for acquisition of targeted $m/z$.

## I. INTRODUCTION

RECTILINEAR or raster scanning typically conducts measurements with statically defined top-down/left-right movements and remains the most common imaging pattern for spectroscopy and microscopy, including Mass Spectrometry Imaging (MSI). MSI measures molecular distributions at high spatial resolutions and chemical specificity. However, raster scanning obtains all of the information within an equipment's Field Of View (FOV), regardless of that information's relevance to research objectives. Whether determining the distribution of a targeted set of molecules or isolating measurements inside the encapsulated tissue area, avoiding the acquisition of non-relevant voxels offers potential for significant throughput gains, even when paired with only basic reconstruction techniques.

Nanospray Desorption Electrospray Ionization (nano-DESI) [1] serves as an example MSI technology, where such approaches may be applied, since it can take hours to days for even a small tissue section to be imaged at high spatial resolutions. Operationally, nano-DESI MSI moves a sample under a dynamic liquid bridge, extracting molecules from its surface and subsequently analyzing these in a mass spectrometer. This process yields the spatial distribution of mass spectra for the sample, quantifying intensities for varying mass-to-charge ($m/z$) channels. Contrary to other popular MSI technologies, particularly Matrix-Assisted Laser Desorption Ionization (MALDI), nano-DESI requires minimal sample preparation and can be conducted outside of a vacuum. However, existing experimental nano-DESI platforms have a line-bounded geometry constraint, performing independent sets of measurements along singular rows. While the movement restriction potentially makes nano-DESI the worst-case MSI technology for dynamic sparse sampling, the throughput can still be notably improved through the integration of DLADS.

### A. Background

Static scanning patterns using predetermined locations can be generated for well-defined and consistent structures [2], alternatively being produced uniformly, randomly, or through stochastic models [3], [4], [5]. A method growing in popularity, within Magnetic Resonance Imaging (MRI) and Computed Tomography (CT), performs random sampling and then relies on deep learning for in-painting. Although a promising prospect for improving imaging throughput, there still exist practical and regulatory requirements for improved result explainability.

An alternative class of algorithms perform dynamic sampling [6], [7], through the merger of stochastic models and compressed sensing, where measurement locations are progressively determined based on data actively being obtained. A particularly successful implementation is the Supervised Learning Approach to Dynamic Sampling (SLADS), first made available in 2017 by Godaliyadda [8]. Thereafter, Scarborough et al. [9] applied SLADS to produce a 20-fold reduction in applied radiation dosage for dynamic X-Ray crystalline protein acquisition, with only a ~0.1% absolute difference compared to a full scan. SLADS, or SLADS-LS, uses a least-squares regression model to produce an Estimated Reduction in Distortion (ERD), indicating the amount of entropy that can be removed from a reconstruction for as-of-yet unmeasured locations, derived from extracted statistical features. SLADS was generally evaluated by Zhang et al. in 2018, reducing the number of required measurements for acceptable reconstructions in confocal Raman microscopy [10]

This project was funded by award UG3HL145593 from the National Institute of Health (NIH) Common Fund, through the Office of Strategic Coordination, as part of the Human BioMolecular Atlas Program's (HuBMAP) Transformative Technology Development division [22].

Code available at github.com/Yatagarasu50469/SLADS version 0.9.2.

D. Helminiak and D. Ye are with the Department of Electrical and Computer Engineering, Marquette University, Milwaukee, WI, 53233 USA (e-mail: david.helminiak@marquette.edu and donghye.ye@marquette.edu).

H. Hu and J. Laskin are with the Chemistry Department, Purdue University, West Lafayette, IN, 47907 USA (e-mail: hu518@purdue.edu and jlaskin@purdue.edu).

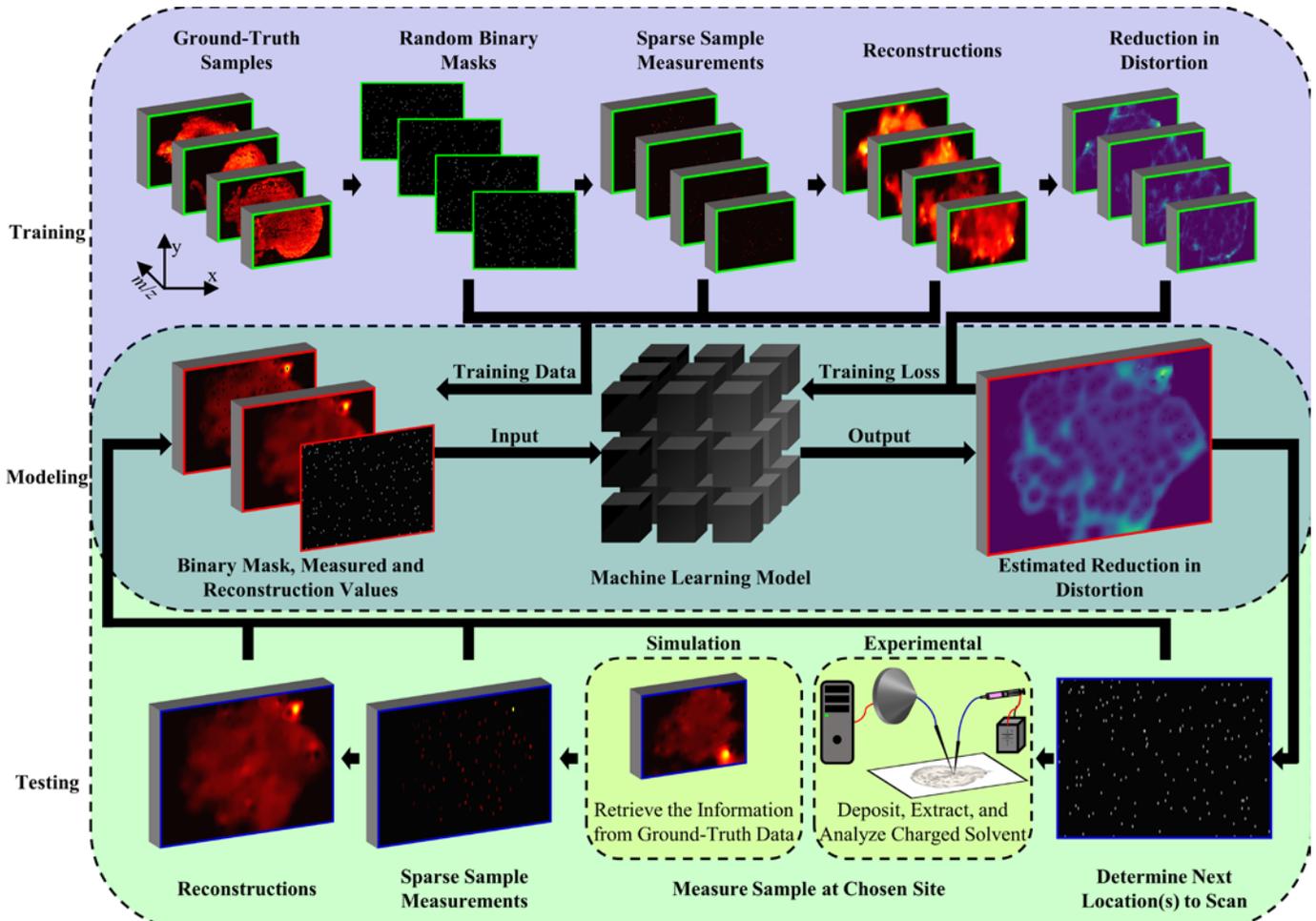

**Fig. 1.** Overview of dynamic sampling models' training and testing procedures. During training, fully measured tissue sample data are sparsely sampled and reconstructed, from which corresponding ground-truth RD are determined. This data trains machine learning models to produce an ERD output, given currently known information, thereby guiding progressive measurements during simulated and experimental acquisitions.

by 6-fold, ~60-95% for Electron Back Scatter Diffraction (EBSD) [11], up to 90% for Energy Dispersive X-Ray Spectroscopy (EDS) [12], and between ~60-90% for metal dendrite sampling when combined with Hierarchical Gaussian Mixture Models (HGMMs) in a model named Unsupervised-SLADS (U-SLADS) [13]. During this time, Zhang et al. also published a multi-model study [14], examining the replacement of the least-squares regression with either a Support Vector Machine (SVM), or a Multi-Layer Perceptron (MLP) network. The MLP, identified as SLADS-Net, improved generalization between training and testing with dissimilar data. An alternate approach was published in 2020 by Grosche et al. [15], using a Probabilistic Approach to Dynamic Image Sampling (PADIS) for single-channel Scanning Electron Microscope (SEM) images, relying on a probability mass function for selection of scanning locations. PADIS outperformed both SLADS-LS and SLADS-Net, which both tended to oversample structural edges.

SLADS has been limited to processing 2-Dimensional (2D) images, where a third dimension (3D) risks obfuscation of channel-specific details [16]. Further, SLADS implementations have focused on structural information, where even if the data was continuous, non-structural areas were homogeneous. MSI data is highly heterogeneous, necessitating more complex models to effectively leverage captured information. An initial study was conducted by Helminiak et al. in 2021 [17], [18] on the feasibility of applying SLADS-LS, SLADS-Net, and a new Deep Learning Approach for Dynamic Sampling (DLADS), using a sequential Convolutional Neural Network (CNN), with nano-DESI MSI tissues. The work reduced the data into 2D by averaging 10 arbitrary mass ranges, empirically observed to contain desirable information. Therein, DLADS reduced the number of required measurements to achieve an acceptable reconstruction for the averaged $m/z$ by 70-80% and demonstrated a 14-46% performance improvement over single-channel SLADS-LS and SLADS-Net. However, where the 2021 study used samples that had already been post-processed. Combination of DLADS with nano-DESI MSI can only be realized after these steps have been re-designed and integrated for actual scans. First, there exists a line-bounded movement constraint on the acquisition probe, where measurements can only be performed along a single indicated line in each scanning cycle. Next, the dimensionality of the measurable locations along that line depend on the actual scan and acquisition rates, which can be variable when using equipment automatic gain control. Additional inconsistencies include the physical start and stop locations for each row, as well as the specific $m/z$ where intensities are measured.

## B. Objectives

This work improves MSI throughput, updating the DLADS algorithm for simulated tissue acquisitions, with principal objectives to 1) incorporate a third dimension of $m/z$ data, 2) optimize the DLADS CNN with a modified U-Net [19], 3) compare DLADS against alternative models, and 4) evaluate theoretical performance of DLADS on real-world tissues, with consideration for experimental platform constraints.

Spatial distributions of different $m/z$ highlight channel-specific structures, whereas integration of common data reduction strategies risks obfuscating fine details, overemphasis on common features, and reduced specificity in model-based location selections. Incorporation of a U-Net architecture improves DLADS ability to extract feature maps from and infer results across a tissue's entire FOV. Quantitative and qualitative assessments are made against a single-channel SLADS-LS, as well as novel multichannel SLADS-LS and SLADS-Net. Further, determined are the effectiveness of considering multiple $m/z$ channels, integrated preprocessing steps, the potential for maximizing scanning throughput, as well as generation of low-error reconstructions for multiple $m/z$. Two operational acquisition modes are presented to accommodate different MSI technologies. Pointwise mode selects future scan positions point-by-point, while a linewise procedure applies an axial constraint, choosing points in a single row each iteration.

## II. METHODS

Fig. 1 shows the procedure for combining dynamic sampling algorithms and models with nano-DESI MSI, divided into stages for training, modeling, and testing. Training extracts data at multiple sampling densities, with random masks applied to a set of fully measured tissues to obtain sets of sparse measurements. Herein, the densities range from 1-30% of the samples' FOV at 1% intervals, replicating the range of information levels that may be encountered during a scan and provide sufficient variance to prevent overfitting during model training. Reconstructions of the measurements for each $m/z$ channel are then compared to the ground-truth data to form Reduction in Distortion (RD) maps. The RD value at a given location represents how much the reconstruction could be improved by scanning that position. Training data passes into the modeling phase, where binary masks, measured, and reconstruction values for each $m/z$ channel constitute a single DLADS input, matched with a single RD map as output. SLADS-LS and SLADS-Net models extract hand-crafted statistical features from the reconstructions for inputs and RD values for unmeasured locations as matched outputs. The resultant models are then queried in testing stage to produce Estimated Reduction in Distortion (ERD) maps for individual $m/z$ channels, where averaging the ERDs allows for the determination and sampling of subsequent location(s).

### A. Reduction in Distortion

Each ground-truth MSI sample $X$ has spatial dimensions of width $m$ and height $n$, wherein $\Omega$ represents the set of all measurable locations, comprised of the subsets $S$ and $T$, containing $k$ measured locations ($S = \{s^{(1)}, s^{(2)}, \ldots, s^{(k)}\}$) and $q$ unmeasured locations ($T = \{t^{(1)}, t^{(2)}, \ldots, t^{(q)}\}$). For consistency in comparing results with prior SLADS implementations, Inverse Distance Weighted (IDW) mean interpolation was employed, using the measured values $X^{(S)}$ and the 10 nearest neighbor distances between $S$ and $T$ to estimate unmeasured location values $\hat{X}^{(T)}$, where a reconstruction $\hat{X}$ comprises $X^{(S)}$ and $\hat{X}^{(T)}$.

DLADS chooses location(s) $P$ to scan from $T$ that minimize error in $\hat{X}$, found through a set of RD values $R$. (1) describes the $R$ of a single $m/z$ channel, as the difference between existing reconstruction error and that formed after scanning each potential location $t$ ($D(\cdot, \cdot)$ being the absolute difference between images). $X$ cannot be known during implementation, where instead ERD values $E$ are determined for unmeasured locations $E^{(T)}$ through trained models, with $E^{(S)} = 0^{(S)}$.

$$R^{(T)} = \left\{ \sum \left( D\left(X, \hat{X}^{(S)}\right) - D\left(X, \hat{X}^{(S \cup t)}\right) \right) : \forall t \in T \right\} \quad (1)$$

During the generation of training and validation sets, as well as evaluation of the testing results, $R$ must be estimated to ensure computational tractability. Determined with (2), for every $t \in T$ a weighted factor $\sigma^{(t)}$ may be found as the minimum distance to a measured location ($s \in S$), divided by a regularization parameter $c$. $R^{(t)}$ for a single $m/z$ channel, may then be approximated through (3) as the sum of a Gaussian filter, with strength $\sigma^{(t)}$ applied to the current reconstruction error. For computational efficiency, the region of filter application and summation may be limited, either by a static or dynamic window. The dynamic option uses a window with radii equal to a multiple of the determined $\sigma^{(t)}$: $\forall t \in T$. Parameter $c$ must be optimized for each intended application of the algorithm from a user defined set $C$. Herein, for each $c \in C$, $R$ informs $P$ in simulated pointwise scanning of the training and validation tissues, where the optimal $c$ maximizes the Area Under the Curve (AUC) value. The AUC may be computed as the integrated Peak Signal to Noise Ratio (PSNR), computed at 1% intervals, and averaged for all reconstructed $m/z$ channels.

$$\sigma^{(t)} = \frac{\min_{s \in S} \|s - t\|}{c} \quad (2)$$

$$R^{(t)} \approx \sum \left( D\left(X, \hat{X}^{(S)}\right) \cdot e^{-\frac{\|s - t\|^2}{2\left(\sigma^{(t)}\right)^2}} \right) \quad (3)$$

$R$ further considers a third-dimension $Z$ of $d$ user-specified $m/z$ ranges, expanded as $R = \{R_{(z)} : \forall z \in Z\}$ and then averaged ($\bar{R} \approx \frac{1}{d} \sum_{z \in Z} R_z$) for sampling location selection. Similarly, the generated ERD expands into $E = \{E_{(z)} : \forall z \in Z\}$, with location selection conducted with the average ERD ($\bar{E} \approx \frac{1}{d} \sum_{z \in Z} E_z$).

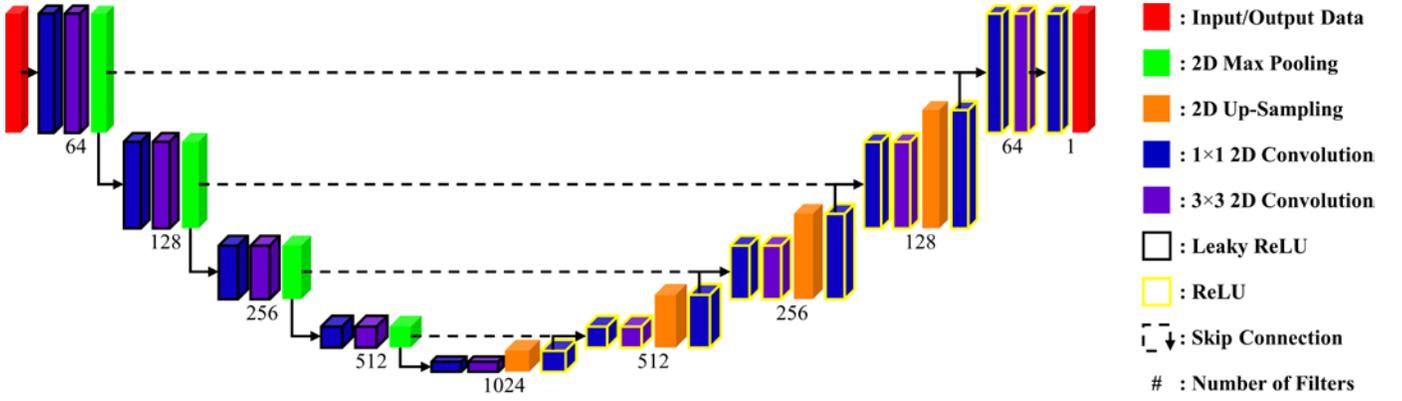

**Fig. 2.** U-Net CNN architecture modified for image-to-image translation, taking in known and estimated $m/z$ information and producing an ERD, which guides the selection of future measurement locations during dynamic sampling acquisitions with DLADS.

*Legend:*
- 🟥 : Input/Output Data
- 🟩 : 2D Max Pooling
- 🟧 : 2D Up-Sampling
- 🟦 : 1×1 2D Convolution
- 🟪 : 3×3 2D Convolution
- ⬜ : Leaky ReLU
- ⬜ (yellow outline) : ReLU
- ⌐ ↕ : Skip Connection
- # : Number of Filters

### B. Models

SLADS-LS and SLADS-Net compute ERD through channel-specific feature vectors $V_{(z)}^{(T)}$, derived from $\hat{X}_z$. Training SLADS-LS produces optimized parameters $\hat{\theta}$, found with (4), where $E_{(z)}^{(T)} = V_{(z)}^{(T)}\hat{\theta}$. SLADS-Net replaces $\hat{\theta}$ with a transformation $g^{(w)}(\cdot)$ (comprising 50 neurons in 5 layers), resulting in $E_{(z)}^{(T)} = g^{(w)}(V_{(z)}^{(T)})$, where network weights $w$ are determined with an Adam solver, Learning Rate (LR) of 1e-3, and minimizing a squared error loss function over 500 epochs.

$$\hat{\theta} = \underset{\theta \in \mathbb{R}}{argmin} \left\| R^{(T)} - V^{(T)}\theta \right\|^2 \tag{4}$$

DLADS updates $g^{(w)}(\cdot)$ with a U-Net architecture (Fig. 2), modified for 3D input(s) to 2D output (enforcing $\bar{E}^{(S)} = 0$ during operation), where being fully convolutional allows for matched input/output spatial dimensions. DLADS uses an input $I$ in (5), containing the reconstruction and measured values, as well as a binary mask of measured positions, each set mapped to their corresponding 2D locations. Two loss functions: Mean Absolute Error (MAE) and Mean-Squared Error (MSE), defined in (6) and (7), are used to train the U-Net weights.

$$I = \left\{ \hat{X}_{(z)}^{(T)}, X_{(z)}^{(S)}, 1^{(S)} \right\} \mapsto \left\{ 0_{m,n}^{(T)}, 0_{m,n}^{(S)}, 0_{m,n}^{(S)} \right\} \tag{5}$$

$$Loss_{MAE} = \left| R_{(z)} - g^{(w)}(I) \right| \tag{6}$$

$$Loss_{MSE} = \left( R_{(z)} - g^{(w)}(I) \right)^2 \tag{7}$$

The network comprises symmetric encoding and decoding halves, first compressing spatial dimensions and doubling the number of filters through back-to-back convolutional layers using Leaky ReLU activations. The decoding section, using ReLU activations, upsamples its inputs with nearest-neighbor interpolation (to avoid visual artifacts [20]), regularly halving the number of filters and re-combining with the encoded feature maps through skip connections. 3×3 kernels are used for extraction of feature maps, while 1×1 kernels are applied for feature map pooling. A minimum of 10 epochs are performed, where termination and restoration of the best-found weights occurs if the validation loss does not improve within 50 epochs.

Data augmentation is used for regularization and enhanced generalization, where after each epoch the initial training set undergoes random horizontal/vertical flips, $\pm45°$ rotation, and height/width/position shifts up to 25%.

### C. Sampling Operation

The implemented pointwise acquisition mode determines the set of points to scan $P$ in (8), as the unmeasured locations with maximal $\bar{E}$. The number of positions may be specified as a percentage of the FOV for group-based acquisition, though this work scans one location per iteration.

$$P = argmax(\bar{E}^{(T)}) \tag{8}$$

There also exists a set $L$ containing the $n$ rows in $X$: $L = \{l^{(1)}, l^{(2)}, \dots, l^{(n)}\}$, with subsets containing only measured $J \subset L = \{l \in L : \exists s \in l\}$ and unmeasured locations $K \subset L = \{l \in L : \nexists s \in l\}$. Formalized in (9) and (10), the linewise acquisition mode first selects the unmeasured row $q$ with the highest sum $\bar{E}$ and then chooses a percentage (with 30% used in this work) of $m$ positions on that line to scan, sorted according to $\bar{E}$.

$$q = argmax\left(\left\{\sum \bar{E}^{(k)} : \forall k \in K\right\}\right) \tag{9}$$

$$P = q^{\left(\underset{\%(m)}{argsort}(\bar{E}^{(q)})\right)} \tag{10}$$

Before engaging dynamic sampling behavior, an initial set of predetermined locations are obtained to provide some information to the models, with a random 1% of the total FOV used for pointwise. Linewise initially scans the same user-defined percentage (30% of positions on pre-selected lines (empirically selected herein as 25%, 50%, and 75% of $n$). The program then enters a loop, using known information as the basis for reconstruction the full sample, and employing the trained models to produce ERD for as of yet unmeasured positions, thereby intelligently guiding measurements until either reaching a user-specified stopping condition, or $\sum \bar{E} = 0$.

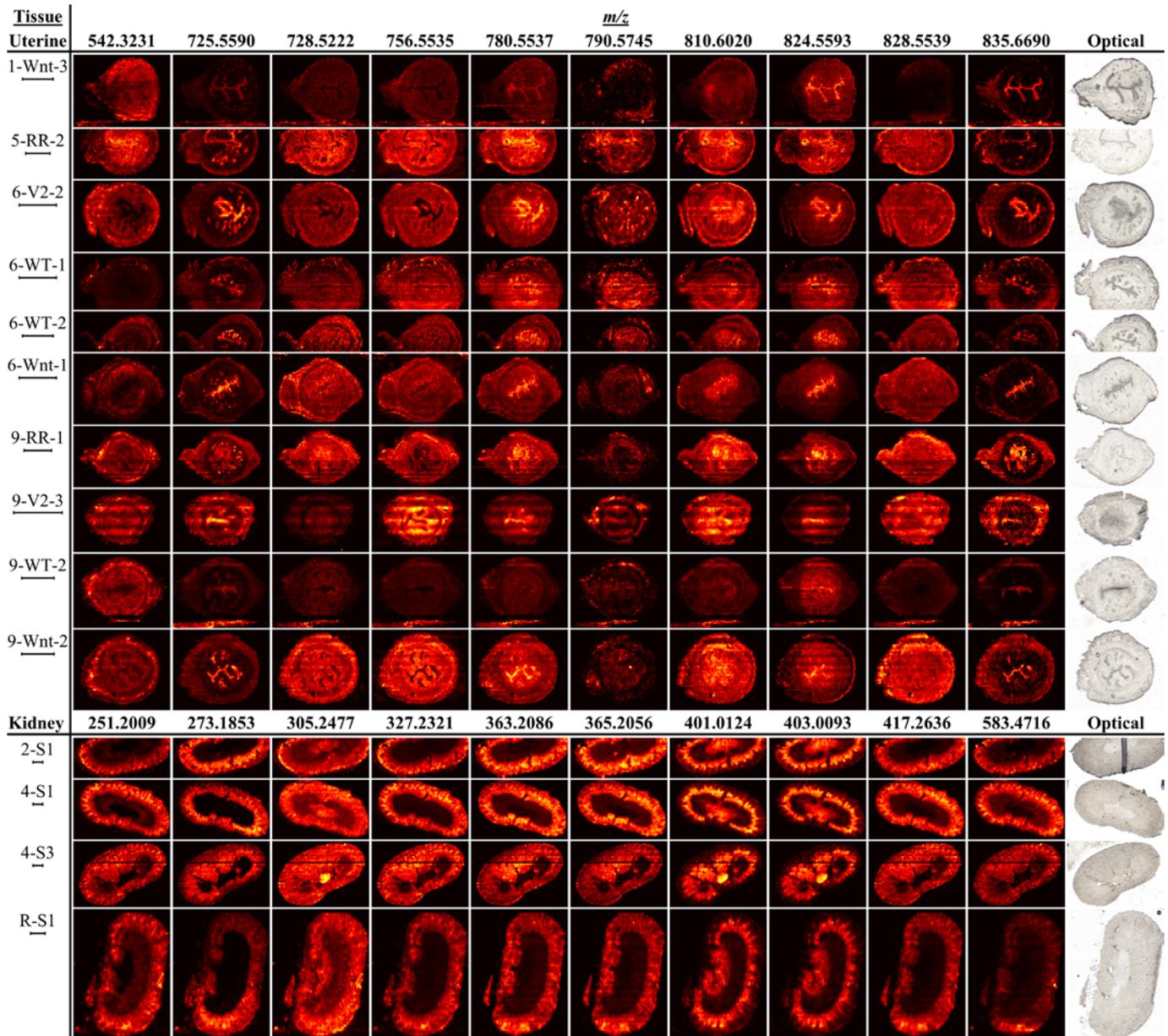

**Fig. 3.** Ground-truth visualizations of 10 expert-selected *m/z*, representative of the desired chemical and biological data. Corresponding optical images, containing visual artifacts from experimental labeling, were registered with MATLAB 2022b *imregister* function for multimodal images, maximizing Mattes mutual information. 1 mm scale bars for each sample, are located beneath the corresponding tissue name.

## III. DATA

Fully acquired nano-DESI MSI data were provided by the Purdue University Department of Chemistry. 10 mouse uterine tissues from a Thermo Fisher Scientific Inc. (Waltham, MA) Q-Exactive HF-X Orbitrap mass spectrometer and 4 mouse kidney samples from an Agilent 6560 IM-QTOF (Santa Clara, CA) system. The uterine samples were randomly split into sets with 6 for training, 2 for validation, and 2 for testing. All 4 kidney tissues were reserved for testing to emphasize generalization capability. The expert-selected, ground-truth *m/z* images $X$ were formed by integrating intensities within a mass range for a peak width ($\Delta$), set at ~20 ppm (parts-per-million). (11) defines the resulting *m/z* range for a central value ($mz$).

$$[mz * (1 - \Delta * 10^{-6}), mz * (1 + \Delta * 10^{-6})] \quad (11)$$

In order to visualize sample geometries, the data must be aligned to a common grid. Typically, this remapping uses a large number of new horizontal positions (e.g., 1,000), between 0 and the maximum horizontal acquisition time across a whole sample, which cannot be known prior to a complete acquisition. Since scanning and acquisition rates can fluctuate, even in the same row, the interpolation to an arbitrary number of values inconsistently changes the physical dimensions each measurement represents. Without consistent distances, IDW mean interpolation creates unpredictable reconstruction artifacts, which are passed into SLADS and DLADS, impacting ERD generation. Given samples where the physical measurement width exceeds the height, SLADS and DLADS have demonstrated tendency to scan positions closer along the vertical axis [17]. These behaviors are corrected by choosing a number of horizontal positions for row realignment equal to the FOV width, divided by the intended equipment scanning rate (μm/s), and multiplied by the intended acquisition rate (spectra/s). Data is temporally re-scaled during distance-dependent computations, forcing voxels to have equivalent physical vertical and horizontal lengths. Fig. 3 visualizes the 10 expert-selected $m/z$, where Table I contains sample metadata (height adjusted for rows removed due to experimental defects).

TABLE I
REPORTED TISSUE AND EXPERIMENTAL METADATA

| Name | Tissue | Set | Width (mm) | Height (mm) | Scan Rate (μm/s) | Acq. Rate (spectra/s) |
|---|---|---|---|---|---|---|
| 1-Wnt3 | Uterine | Testing | 3.1 | 2.304 | 15.0 | 1.0 |
| 5-RR-2 | Uterine | Training | 3.9 | 1.995 | 15.0 | 1.0 |
| 6-V2-2 | Uterine | Testing | 2.6 | 1.922 | 10.0 | 1.0 |
| 6-WT-1 | Uterine | Training | 2.6 | 1.530 | 10.0 | 1.0 |
| 6-WT-2 | Uterine | Training | 3.2 | 1.333 | 10.0 | 1.0 |
| 6-Wnt-1 | Uterine | Training | 2.6 | 1.920 | 10.0 | 1.0 |
| 9-RR-1 | Uterine | Training | 3.6 | 2.275 | 15.0 | 1.0 |
| 9-V2-3 | Uterine | Training | 2.0 | 1.290 | 10.0 | 1.0 |
| 9-WT-2 | Uterine | Validation | 3.0 | 2.277 | 15.0 | 1.0 |
| 9-Wnt-2 | Uterine | Validation | 3.0 | 2.511 | 15.0 | 1.0 |
| 2-S1 | Kidney | Testing | 9.501 | 4.0 | 40.2 | 1.0 |
| 4-S1 | Kidney | Testing | 9.999 | 6.2 | 40.2 | 1.0 |
| 4-S3 | Kidney | Testing | 9.799 | 6.66 | 40.2 | 1.0 |
| R-S1 | Kidney | Testing | 6.7 | 9 | 40.2 | 1.0 |

## V. EXPERIMENTS

Simulations were conducted on an AMD Ryzen 3970X platform, 128 GB DDR4 RAM, 3 NVIDIA RTX 2080 Ti GPUs, and Ubuntu 20.04. Results were evaluated with: 1) the average (avg.) $m/z$ PSNR AUC, found as the integrated average PSNR of the $m/z$ reconstructions over the FOV sampling densities, 2) the average ERD PSNR AUC, where a comparative RD can be calculated after simulation, and 3) the best epoch, where validation loss during training was minimized. The average $m/z$ PSNR AUC indicates the practical effects of using a given model, while the average ERD PSNR AUC quantifies the actual inter-model capabilities. The best epoch serves as a reference for overall training efficiency. The termination criteria for scanning includes reaching ~30% measured FOV for pointwise, every row is visited once in linewise, or when the $\bar{E} = 0$.

### A. Parameter Optimization

A set of potential values for the regularization parameter $c$: $C = \{1, 2, 4, 8, 16, 32, 64, 128, 256\}$, were used to consider use of a single (arbitrarily chosen as 542.3231) and multiple (all expert-selected) $m/z$ channels, as well as static (11×11 to 19×19) and dynamic (1σ to 5σ) windows for $\bar{R}$ generation. Simulated pointwise scans using $\bar{R}$ were performed on the training and validation sets, up to ~30% measured FOV. Resulting metrics, for a static 15×15 window (Table II and Fig. 4), demonstrate that using multiple $m/z$ provides an average 4.7% AUC advantage across all $C$. The optimal $c$ was found for a single $m/z$ to be 4, but 8 with multiple. Employing multiple $m/z$ and $c = 8$ (Table III and Fig. 5), shows the dynamic 3σ window yielded the best AUC, taking 0.11 s less to compute $\bar{R}$, compared to the best performing static window (15×15).

TABLE II
AUC PERFORMANCE USING $\bar{R}$ ACROSS TRAINING AND VALIDATION SETS WITH A 15×15 STATIC WINDOW

| | Single $m/z$ | | Multiple $m/z$ | |
|---|---|---|---|---|
| $c$ | Avg. $m/z$ PSNR AUC | Avg. RD Compute Time (s) | Avg. $m/z$ PSNR AUC | Avg. RD Compute Time (s) |
| 1 | 795.52 | 0.59 | 809.59 | 0.59 |
| 2 | 798.71 | 0.59 | 818.98 | 0.59 |
| 4 | 800.56 | 0.59 | 834.71 | 0.59 |
| 8 | 789.52 | 0.59 | 838.20 | 0.59 |
| 16 | 772.41 | 0.60 | 818.91 | 0.59 |
| 32 | 758.13 | 0.60 | 802.99 | 0.59 |
| 64 | 757.12 | 0.60 | 799.85 | 0.60 |
| 128 | 756.51 | 0.55 | 799.76 | 0.55 |
| 256 | 756.51 | 0.41 | 799.76 | 0.42 |

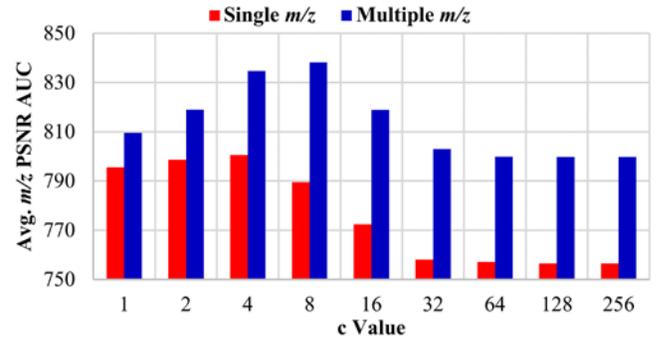

**Fig. 4.** Averaged AUC scores for the training and validation sets, yielded across potential $c$ values, using $\bar{R}$ produced with a static 15×15 window, for both single and multiple $m/z$

TABLE III
AUC PERFORMANCE USING $\bar{R}$ ACROSS TRAINING AND VALIDATION SETS WITH MULTIPLE $m/z$ AND $c = 8$

| Window | Size | Avg. $m/z$ PSNR AUC | Avg. RD Compute Time (s) |
|---|---|---|---|
| Dynamic | 1σ | 838.16 | 0.28 |
| | 2σ | 838.15 | 0.28 |
| | 3σ | 838.25 | 0.28 |
| | 4σ | 838.20 | 0.29 |
| | 5σ | 838.20 | 0.30 |
| Static | 11×11 | 838.14 | 0.35 |
| | 13×13 | 838.06 | 0.37 |
| | 15×15 | 838.20 | 0.39 |
| | 17×17 | 838.20 | 0.43 |
| | 19×19 | 838.20 | 0.46 |

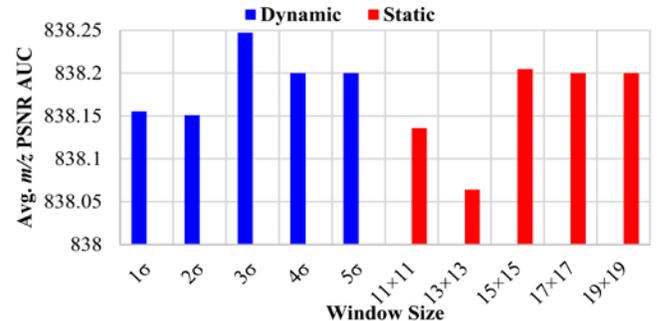

**Fig. 5.** Averaged AUC scores for the training and validation sets, produced using $\bar{R}$ from static and dynamic windows.

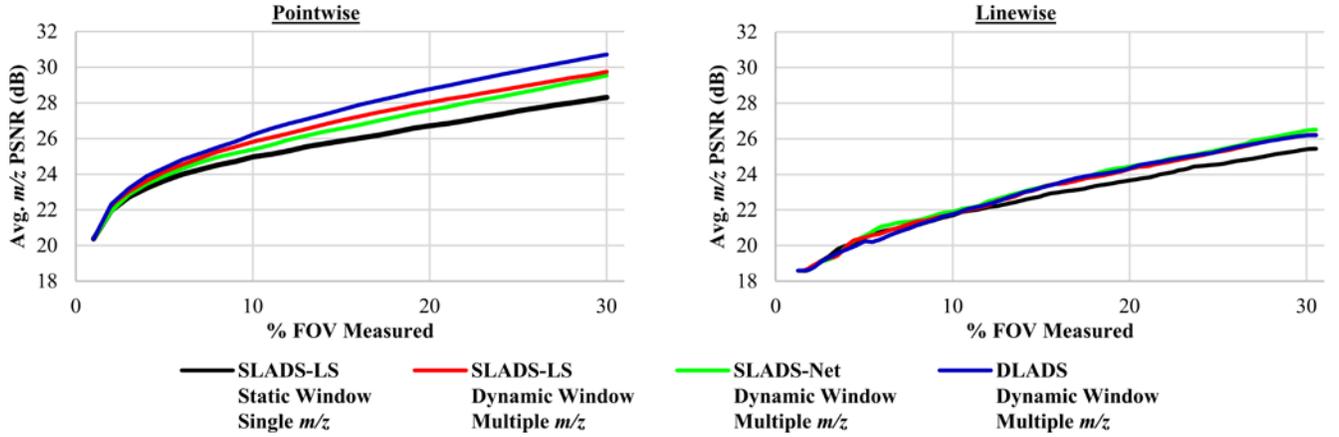

**Fig. 6.** Averaged PSNR of the *m/z* reconstructions in testing across the percent FOV measured for pointwise and linewise acquisition modes.

TABLE V
QUANTIFIED MEASURES FOR SIMULATED SCANNING OF THE TESTING DATA

| Acquisition Mode | *m/z* Input(s) | Window | Model | Final Avg. % Measured | Final Avg. *m/z* PSNR (dB) | Avg. *m/z* PSNR AUC | Avg. ERD PSNR AUC | Avg. ERD Compute Time (s) |
|---|---|---|---|---|---|---|---|---|
| Pointwise | Single | Static | SLADS-LS | 30.00 | 28.31 | 745.46 | 536.93 | 0.23 |
| | Multiple | Dynamic | SLADS-LS | 30.00 | 29.75 | 775.18 | 725.75 | 0.30 |
| | | | SLADS-Net | 30.00 | 29.54 | 765.24 | 731.34 | 0.54 |
| | | | DLADS | 30.00 | 30.71 | 791.95 | 778.42 | 0.25 |
| Linewise | Single | Static | SLADS-LS | 30.25 | 25.44 | 664.96 | 610.97 | 0.30 |
| | Multiple | Dynamic | SLADS-LS | 30.25 | 26.22 | 676.21 | 744.03 | 0.39 |
| | | | SLADS-Net | 30.25 | 26.51 | 680.04 | 688.30 | 0.63 |
| | | | DLADS | 30.25 | 25.98 | 675.98 | 731.20 | 0.27 |

## B. DLADS Hyperparameters

Different hyperparameter combinations were attempted with DLADS including: 1) MAE and MSE losses, 2) Adam, Nadam, and RMS (Root Mean Squared) optimizers, as well as 3) LRs of 1e-3, 1e-4, 1e-5, and 1e-6. Successful training results are presented in Table IV. A MSE loss, RMS optimizer, and LR of 1e-4 produced the highest average ERD PSNR, relative to the ground-truth RD. However, for minimizing the overall *m/z* reconstruction error, the optimal hyperparameters were found to be a Nadam optimizer, LR of 1e-4, and MAE loss.

TABLE IV
QUANTIFIED MEASURES WITH THE VALIDATION SET
OBTAINED BY VARYING DLADS HYPERPARAMETERS

| Loss | LR | Optimizer | Best Epoch | Avg. *m/z* PSNR AUC | Avg. ERD PSNR AUC |
|---|---|---|---|---|---|
| MAE | 1e-3 | RMS | 106 | 843.19 | 734.44 |
| | 1e-4 | Adam | 115 | 842.16 | 730.39 |
| | | Nadam | 325 | 844.93 | 732.22 |
| | | RMS | 97 | 842.74 | 709.24 |
| | 1e-5 | Adam | 134 | 841.23 | 730.16 |
| | | Nadam | 324 | 844.36 | 740.58 |
| | | RMS | 168 | 840.25 | 725.43 |
| | 1e-6 | Adam | 584 | 836.76 | 699.79 |
| | | Nadam | 584 | 836.97 | 694.83 |
| | | RMS | 328 | 838.94 | 695.58 |
| MSE | 1e-4 | Nadam | 34 | 776.04 | 666.53 |
| | | RMS | 109 | 840.44 | 784.15 |
| | 1e-5 | Adam | 95 | 827.28 | 690.70 |
| | | Nadam | 164 | 829.22 | 689.48 |
| | | RMS | 176 | 829.35 | 696.07 |
| | 1e-6 | Adam | 454 | 820.76 | 639.70 |
| | | Nadam | 610 | 823.02 | 644.75 |
| | | RMS | 506 | 823.98 | 680.52 |

## C. Simulation Testing

Final evaluations were conducted with the testing dataset for both pointwise and linewise acquisition modes. SLADS-LS, SLADS-Net, and DLADS were trained on RD generated with multiple *m/z* channels, $c = 8$, and a $3\sigma$ dynamic window. A comparative version of SLADS-LS, using only a single *m/z* channel (arbitrarily chosen as 542.3231 for uterine and 251.2009 for kidney tissues) represents a comparative baseline, trained with RD generated using $c = 4$ and a static 15×15 window. Fig. 6 shows the averaged PSNR of the *m/z* reconstructions (Table V), as plotted against the percent FOV measured. Progressive binary measurement masks for testing sample 1-Wnt-3 are shown in Fig. 7, with average PSNR of the *m/z* reconstructions tabulated in Table VI. Corresponding progressive *m/z* reconstructions for the DLADS model are visualized in Fig. 8, where Table VII holds the PSNR scores.

TABLE VI
AVERAGE PSNR (dB) OF *m/z* RECONSTRUCTIONS FOR
SIMULATED SCANNING OF TESTING SAMPLE 1-WNT-3

| *m/z* Inputs, Window | Model | Pointwise: ~% FOV | | | Linewise: % Lines | | |
|---|---|---|---|---|---|---|---|
| | | 10 | 20 | 30 | 25 | 50 | 100 |
| Single, Static | SLADS-LS | 27.72 | 29.09 | 30.28 | 25.07 | 26.10 | 27.95 |
| Multiple, Dynamic | SLADS-LS | 28.77 | 30.77 | 32.29 | 25.43 | 26.84 | 28.47 |
| | SLADS-Net | 28.48 | 30.55 | 32.36 | 25.44 | 26.74 | 28.32 |
| | DLADS | 29.15 | 31.52 | 33.28 | 25.38 | 26.87 | 28.44 |

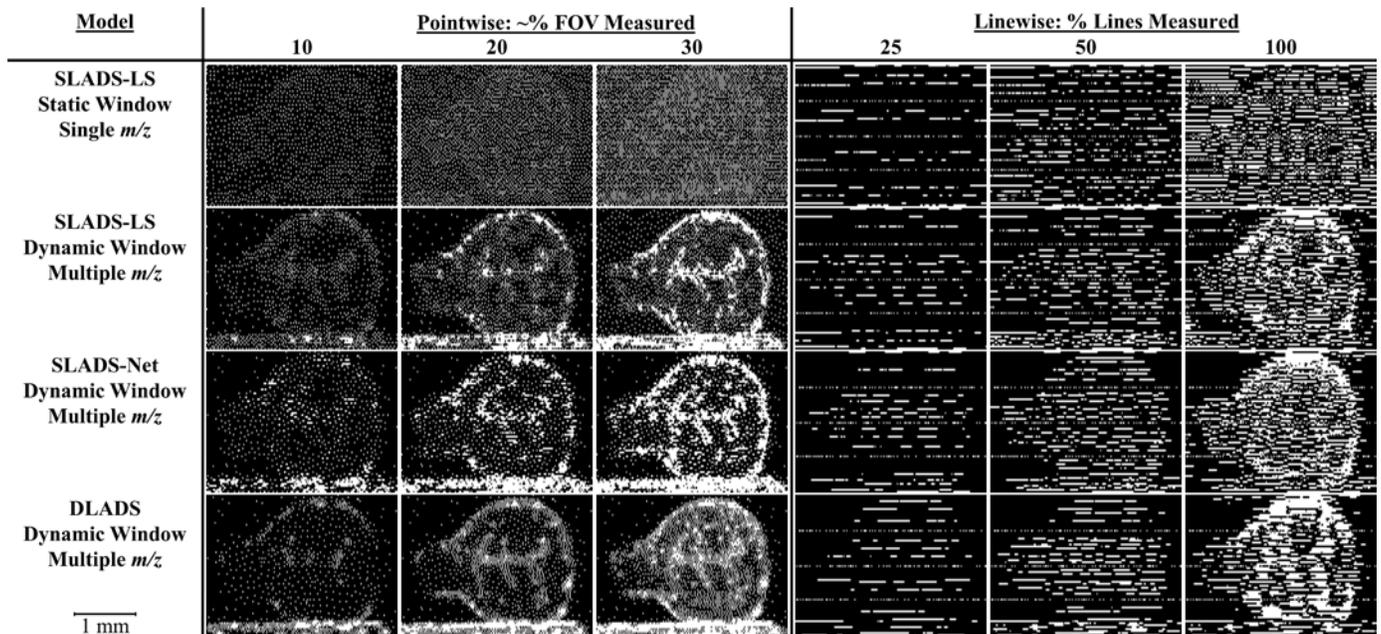

**Fig. 7.** Measurement masks from simulated scanning of testing sample 1-Wnt-3 for pointwise and linewise acquisition modes.

TABLE VII
AVERAGE PSNR OF *m/z* RECONSTRUCTIONS FOR DLADS
SIMULATED SCANNING OF TESTING SAMPLE 1-WNT-3

| | Avg. *m/z* PSNR (dB) | | | | | |
|---|---|---|---|---|---|---|
| | Pointwise: ~% FOV | | | Linewise: ~% Lines | | |
| *m/z* | 10 | 20 | 30 | 25 | 50 | 100 |
| 542.3231 | 25.46 | 27.26 | 29.16 | 26.39 | 27.86 | 29.06 |
| 725.5590 | 33.17 | 36.10 | 38.71 | 28.45 | 30.12 | 31.29 |
| 728.5222 | 29.91 | 32.12 | 33.51 | 25.77 | 27.01 | 27.95 |
| 756.5535 | 31.25 | 33.90 | 35.57 | 25.85 | 27.34 | 28.45 |
| 780.5537 | 29.75 | 32.59 | 34.42 | 25.51 | 26.91 | 28.35 |
| 790.5745 | 24.08 | 25.68 | 27.30 | 22.90 | 23.89 | 25.86 |
| 810.6020 | 31.08 | 33.89 | 35.71 | 27.23 | 28.89 | 31.33 |
| 824.5593 | 25.49 | 27.78 | 29.49 | 23.42 | 25.07 | 27.51 |
| 828.5539 | 34.89 | 36.92 | 38.51 | 29.09 | 30.97 | 31.76 |
| 835.6690 | 26.41 | 28.92 | 30.43 | 22.58 | 24.35 | 25.38 |

During pointwise acquisition, compared to the single *m/z* SLADS-LS, multiple *m/z* SLADS-LS, and multiple *m/z* SLADS-Net, DLADS achieved boosts of 6.0%, 2.1%, and 3.4% to the average *m/z* PSNR AUC, as well as 36.7%, 7.0%, and 6.2% to the average *m/z* ERD AUC. The three models with multiple *m/z*, demonstrated an average 1.8% improvement over the single *m/z* SLADS-LS and all achieved roughly equivalent *m/z* reconstruction performance, with SLADS-Net having the best final average *m/z* PSNR, but the multiple *m/z* SLADS-LS having the best average ERD PSNR AUC. While DLADS computation of the average ERD using multiple *m/z* inputs was faster by ~18% for pointwise and ~36% for linewise compared to the updated SLADS-LS model, all models demonstrated they add negligible additional temporal overhead to MSI procedures.

The visualized binary masks (Fig. 7) illustrate the tendency of SLADS models to oversample structural edges. Incorporation of spatial relationships in DLADS reduces this behavior through convolutional layers and abstraction of extracted feature maps using the U-Net architecture, seen in its global selection of diverse measurement locations, yet retained focus on tissue structures. The addition of the linewise axial constraint conflicts with the principal axioms of compressed sensing, where effective reconstruction depends on having sufficient sparsity in the obtained signal for a global context. Since PSNR evaluates the whole of a reconstructed *m/z* image, rather than just desired tissue content, this has an appreciable impact on the quantitative metrics. Where the SLADS models capture more tissue edge data, there are cleaner boundaries in the generated reconstructions and thereby less overall reconstruction error. This behavior can be seen in linewise acquisition, where the sampling pattern for DLADS has prioritized internal structures, but achieved a lower *m/z* reconstruction score compared to SLADS-Net, which had greater variance in its chosen sampling locations.

## V. FUTURE WORK

Since this study was conducted, an earlier version of DLADS underwent a proof-of-concept integration with an experimental nano-DESI MSI platform demonstrating a 2.3-fold throughput improvement with more restrictive linewise constraints [22]. Investigation into DLADS potential for alternate imaging technologies, particularly MALDI MSI, given its inherent pointwise capability, are planned avenues for further studies. A major limitation of this work was its evaluation of only targeted *m/z* channel reconstructions. Considering reconstruction across the whole MSI spectrum would allow for understanding how well targeted *m/z* represent the whole data and offer a potential direction for researchers without specific molecules of interest in mind. Additional work should include automatic stopping criteria for scanning, testing of the group-based pointwise acquisition mode, dynamic selection of pertinent *m/z* and more varied probe movement constraints. A wider cross-tissue study would also be desirable to evaluate DLADS generalization capability. Finally, the performance gains in ERD quality, in transitioning from an MLP to a CNN, indicates potential worth in incorporating adversarial neural networks for both ERD determination and *m/z* reconstructions.

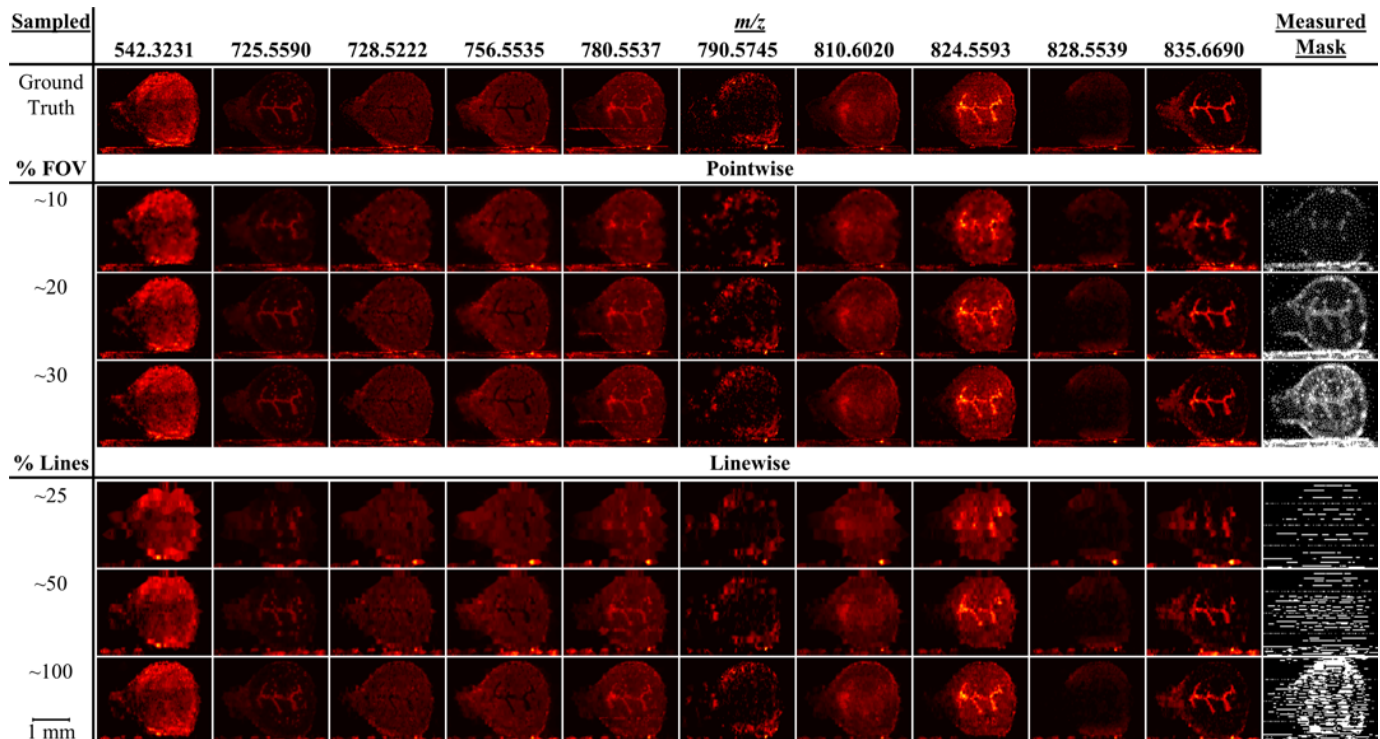

**Fig. 8.** *m/z* reconstructions for simulated scanning of sample 1-Wnt-3 with DLADS for pointwise and linewise acquisition modes.

## VI. CONCLUSION

Where prior SLADS and dynamic sampling studies have focused on structural content, the consideration of increased dimensionality illustrates the potential for DLADS to be utilized for technologies involving heterogeneous content. Herein, an improved version of DLADS further advances the dynamic sampling methodology for integration with MSI, through the incorporation of a third dimension for mass ranges, updated RD generation procedure, consideration for existing physical hardware constraints, and negligible additional temporal overhead. All multiple *m/z* models demonstrated increased performance both in terms of their regression capabilities and produced reconstruction quality, with DLADS achieving respective advancements of 36.7% and 6.0%. The simulated integration of updated SLADS and DLADS dynamic sparse sampling algorithms provide a 70% minimum throughput improvement to nano-DESI MSI for targeted *m/z*.

**David Helminiak** (Member, IEEE), David Helminiak is a Research Assistant and Ph.D. candidate in the Electrical and Computer Engineering department at Marquette University, where he received an M.S. (2021) and two B.S. (2018) degrees. His research focuses on developing neural network architectures for interdisciplinary applications of artificial intelligence and machine learning.

**Hang Hu**, Hang Hu studies mass spectrometry imaging with Prof. Julia Laskin. He received his Ph.D. (2022) from Purdue University, M.E. from Fudan University (2017), and B.Sc. degree from East China University of Science and Technology (2015). His research interests include development of automated systems for high-throughput mass spectrometry imaging experiments and computational methods for analyzing mass spectrometry imaging data.

**Julia Laskin**, Julia Laskin is the William F. and Patty J. Miller Professor of Analytical Chemistry at Purdue University. She obtained her Ph.D. (1998) from The Hebrew University of Jerusalem and M.S. (1990) from Leningrad Polytechnic Institute. Her research focuses on understanding of physical and chemical phenomena underlying molecular imaging of biological samples and chemical analysis of complex mixtures.

**Dong Hye Ye** (Member, IEEE), Dong Hye Ye is an Assistant Professor in the Electrical and Computer Engineering department at Marquette University. He received his Ph.D. (2013) from the University of Pennsylvania, M.S. (2008) from the Georgia Institute of Technology, and B.S. (2007) at Seoul National University. His research includes sparse sampling for microscopic imaging, iterative tomographic reconstruction, and UAV sensing via machine learning.